# Theoretical method for calculation of effective properties of composite materials with reconfigurable microstructure: electric and magnetic phenomena

Andrei A. Snarskii[1,2], Denis Zorinets[1], Mikhail Shamonin[3*], Viktor M. Kalita[1,4]

[1] *National Technical University of Ukraine "Igor Sikorsky Kyiv Polytechnic Institute", Prospekt Peremohy 37, 03056 Kiev, Ukraine*

[2] *Institute for Information Recording, NAS of Ukraine, Shpaka Street 2, 03113 Kiev, Ukraine*

[3] *East Bavarian Centre for Intelligent Materials (EBACIM), Ostbayerische Technische Hochschule (OTH) Regensburg, Seybothstr. 2, 93053 Regensburg, Germany*

[4] *Institute of Physics, NAS of Ukraine, Prospekt Nauky 46, 03028 Kiev, Ukraine*

**ABSTRACT**

We propose a theoretical approach for calculating effective electric and magnetic properties of composites with field-dependent restructuring of the filler. The theory combines the Bruggeman-Landauer approximation extended to a field-dependent (variable) percolation threshold with the approximate treatment of nonlinearity of material properties. Theoretical results are compared with experiments on magnetorheological elastomers, which in the context of investigated phenomena are often called magnetoactive elastomers (MAEs). In MAEs with soft polymer matrices, the mutual arrangement of inclusions changes in an applied magnetic field. This reorganization of the microstructure leads to unconventionally large changes of electrical and magnetic properties. Obtained theoretical results describe observed phenomena in MAEs well. Qualitative agreement between theory and experiment is demonstrated for the magnetodielectric effect. In the case of magnetic permeability, quantitative agreement is achieved. The theoretical approach presented can be useful for development of field-controlled smart materials and design of smart structures on their basis, because the field dependence of physical properties can be predicted.

---

[*] Corresponding author. E-Mail: mikhail.chamonine@oth-regensburg.de

# I. Introduction

By composite material, we mean a macroscopically inhomogeneous medium consisting of two or more "phases", where each phase is a material with distinct physical properties. The structure is assumed to be random heterogeneous, for example, randomly located inclusions of the first phase dispersed in a continuous matrix (second phase). There may be many different modifications or variations of the microstructure, e.g. tiny spheres in the matrix, anisotropic inclusions, elongated inclusions with a random arrangement of their centers and/or a random arrangement of axes, and the like. Many real-world composites can reasonably be represented by those structures (after neglecting fine details). On a large spatial scale, which should be much bigger that the characteristic size of the correlation length $\xi$ [1-4], the composite is homogeneous, one speaks of the averaged homogeneity. Effective coefficients describe physical properties of such a composite material. Theoretical description of the effective properties of composites with macroscopic inclusions boils down to the calculation of effective coefficients, e.g. magnetic permeability. For instance, in the case of a macroscopically inhomogeneous magnetic composite with different values of the magnetic permeability of the phases, the effective magnetic permeability relates the magnetic field strength to the magnetic flux density, where both physical quantities are averaged over the volume.

In all the above examples, including those given in cited seminal books, the microstructure of the composites fixed and various physical processes take place at this given structure [2-6].

In the present paper, we consider an alternative statement of the problem, when the microstructure (e.g. mutual arrangement of filling particles) of a composite material changes under the influence of external fields, also in the situation when dimensions and the shape of the sample remain constant.

In this case, one often says that a restructuring (RS) of the filler [7] or particles [8,9] takes place. RS can also happen during measurement. In the following, we will use as a striking example magnetoactive elastomers (MAEs) at room temperature [10,11]. For instance, when measuring the effective magnetic permeability of a MAE with a soft polymer matrix, it is necessary to put a MAE specimen into a magnetic field. In an inhomogeneous medium, the magnetic field will be non-uniform and may lead to a displacement of inclusions, i.e. to RS of the particulate filler.

Magnetoactive elastomers are composite materials where μm-sized ferromagnetic filler particles are in a soft elastomer matrix [12-15]. In an external magnetic field, these particles change their arrangement within the composite material (i.e. RS of the particulate filling takes place). It is assumed that the external dimensions of a sample don't change, that requires specific experimental conditions. It is important that the matrix is bound to the inclusions. When an external magnetic field vanishes, elasticity of the matrix forces the initial distribution of particles to recover. Any physical property related to the arrangement of inclusions within the composite material will change [16]. If the matrix is soft (Young's modulus is smaller than approximately 30 kPa), the RS of the particulate filler can be significant even at moderate (100 - 300 mT) magnetic flux densities [7,11,17-21]. One of the most known appearances of such a RS is the magnetorheological (MR) or field-stiffening effect, where the relative changes of the elastic moduli can reach three to four orders of magnitude in saturating magnetic fields [22,23]. Hitherto, the majority of research has concentrated on the MR effect because of its industrial applications [24-26]. Similar RS effects were also observed in the electrorheological elastomers where the arrangement of dielectric particles can be influenced by external electric fields [27-29]. A more exotic effect is the recently reported rotation of hard inclusions in a soft matrix during mechanical loading [30].

Initially, it is assumed that there are material equations corresponding to the physical phenomenon under consideration, for example, $\mathbf{B} = \mu_0 \mu \mathbf{H}$, $\mathbf{j} = \sigma \mathbf{E}$, $\mathbf{D} = \varepsilon_0 \varepsilon \mathbf{E}$, where $\mathbf{B}$, $\mathbf{j}$, and $\mathbf{D}$ are the magnetic flux density, the electric current density and the electric displacement field, respectively, while $\mathbf{H}$ is the magnetic field strength and $\mathbf{E}$ is the electric field strength. $\mu_0$ is the magnetic permeability of vacuum, and $\varepsilon_0$ is the vacuum permittivity. In the quasi-static approximation [31], equations $div\mathbf{B} = 0$, $div\mathbf{j} = 0$, $div\mathbf{D} = 0$, and $curl\mathbf{H} = 0$, $curl\mathbf{E} = 0$ are satisfied. Of course, all these vectors and local coefficients (magnetic permeability $\mu$, specific electrical conductivity $\sigma$, permittivity, sometimes called dielectric constant, $\varepsilon$) depend on the coordinates **r**. The effective coefficients, namely the effective relative magnetic permeability $\mu^e$ and the effective relative dielectric permittivity $\varepsilon^e$ (dielectric constant) are defined as

$$\langle \mathbf{B} \rangle = \mu_0 \mu^e \langle \mathbf{H} \rangle, \qquad \langle \mathbf{D} \rangle = \varepsilon_0 \varepsilon^e \langle \mathbf{E} \rangle, \tag{1}$$

where $\langle ... \rangle = 1/V \int ... dV$, $V$ is the averaging volume, wherein the characteristic dimensions of the averaging region should be much larger than the correlation length $\xi$.

Many examples of different structures of composites and methods for calculating effective coefficients are given in books (see [4,5] and others). It is obvious that the exact solution of the problem of calculating the effective coefficients is possible only in a few exceptional cases [32-35]. Therefore, one has to use various approximate methods.

It is obvious that conventional methods for calculating effective properties that do not take into account RS in a specimen placed into an external field are not applicable to description of the properties of MAEs and in other similar cases. We propose an approximate approach for describing the effective properties of composites with RS based on the concept of a movable percolation threshold.

The paper is organized as follows: In Section II, we present a qualitative description of RS phenomena and formulate the main idea behind the proposed theoretical approach. Here we will use the model of percolation structure, when up to the percolation threshold (not necessarily in the critical region, that is, just close to the percolation threshold) in a macroscopically disordered system, there are finite clusters forming the so-called pre-cluster. With an increase in the concentration of inclusions, the pre-cluster will become a part of an infinite cluster, forming a continuous path via the inclusions through the entire sample. In Section III, we go beyond the conventional effective-medium theory (EMT) and describe a modification of it allowing one to "tune the percolation threshold". In Section IV, we use the approach discussed in the preceding section for description of the magnetocapacitive effect in MAEs. Section V presents the treatment of nonlinear properties of the filler for calculating the effective magnetic permeability in the absence of RS. Theoretical methods outlined in Section

III and V are unified into a coherent approach in Section VI, where this approach is successfully applied to calculation of nonlinear magnetic permeability in MAEs with RS of the filler. The physical implications of the proposed theoretical approach are discussed in Section VII. Conclusions are drawn in Section VIII. Two Appendixes provide details on definition of percolation threshold in the Bruggeman-Landauer (BL) approximation and nonlinear equations for effective magnetic permeability.

**II. Qualitative description of restructuring**

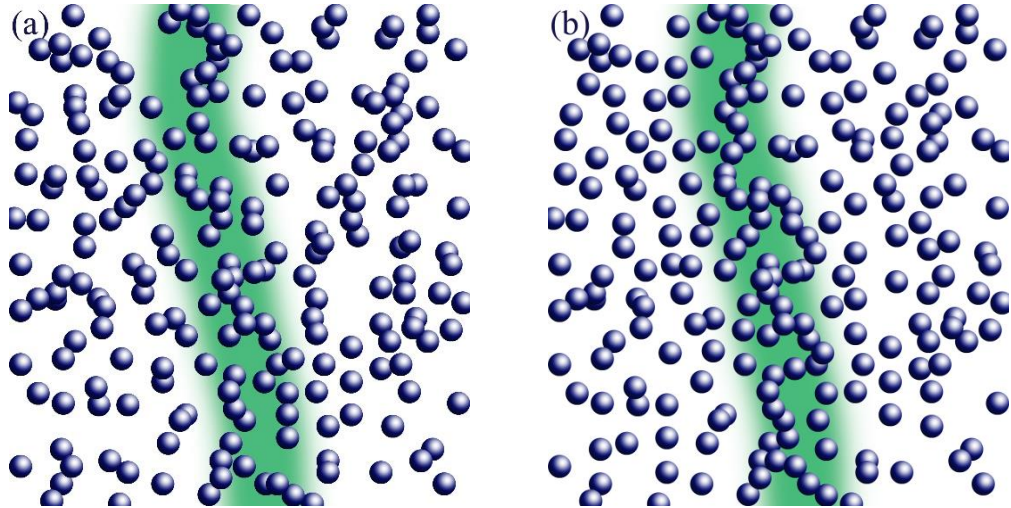

**Figure 1.** (a) Qualitative representation of the microstructure of a composite at $p < p_c$. (b) Qualitative representation of the microstructure of a composite at $p > p_c$. A colored stripe designates the region corresponding to the pre-cluster. The pre-cluster is transformed into an infinite cluster shown as a colored stripe.

Let us qualitatively consider the RS phenomenon on the example of an MAE. The distribution of the magnetic field strength and the magnetic field strength inside a composite is homogeneous only on average. For example, on the dimensions of the order of the correlation length below the percolation threshold (when the concentration of inclusions is less than the percolation threshold), the composite structure is a set of finite clusters of inclusions. The size of these clusters is less than the correlation length $\xi$. With an increase in the concentration of inclusions, the average size of these finite clusters grows. At a certain concentration value, which is the percolation threshold $p_c$, the finite clusters (but not all) connect and form the so-called infinite cluster passing through the entire specimen. This infinite cluster is a random network with a characteristic cell size $\xi$. Inside this network, without touching it, there are finite clusters that have not joined to the network yet. Thus, when the concentration is less than the threshold value $p < p_c$, in the composite there is an assortment of finite clusters (we will call it a pre-cluster), which, with an increase in the concentration of inclusions $p \to p_c$, will connect parts of the PC and form an infinite cluster, see Figure 1 as an illustration.

Consider the situation when a magnetic field is applied. We assume that the elastomer matrix is non-magnetic, i.e. the relative magnetic permeability $\mu_2 = 1$, and inclusions consisting of carbonyl iron with dimensions of the order of a micrometer have relative magnetic permeability $\mu_1 > \mu_2$, and in some cases $\mu_1 \gg \mu_2$. The closer is the concentration of inclusions $p$ to $p_c$, i.e. the denser is the pre-cluster (the relative number of inclusions in pre-cluster, normalized to the total number of particles, is increased) the greater part of the magnetic flux will pass through the pre-cluster. Accordingly, in a pre-cluster, the magnetic flux density $\mathbf{B}(\mathbf{r})$ is larger in the pre-cluster than outside of it. A magnetic field gradient comes into play and the force acts on magnetic inclusions in vicinity of the pre-cluster. This force is proportional to $grad(H^2)$ [31] and is directed towards the pre-cluster. At room temperature, the elastic matrix is soft and the particles with the attached matrix can move within the specimen (until the elastic force from the matrix stops them) and thereby increase the relative number of particles (i.e. their concentration) in the pre-cluster. We can say that the pre-cluster grows. Despite the fact that the total number of particles in the sample is constant, after displacement of inclusions the compacted pre-cluster looks as if the particles did not change their locations, but the concentration of inclusions in the sample has been increased. A visualization of such a process was carried out in [36,37], cf. Fig. 5 (b) in [37]. The attractive forces between magnetic particles were sufficiently strong to bring the particles close to contact in spite of counteracting elastic forces from a deformed polymer matrix. Formation of chain-like structures under the influence of an external magnetic field of 270 mT is clearly seen in Fig. 5 (b) of [37].

The main idea of the proposed theoretical description of effective properties in the case of the re-arrangement of inclusions is that the increase of their concentration in the pre-cluster can be interpreted as a decrease in the percolation threshold. This means that the reduction of the difference $p_c - p$ is not attributed to a (local) increase in $p$, but to a decrease in $p_c$. With such a description, the percolation threshold is no longer a constant, but is a function of the magnetic field that decreases with increasing $\langle \mathbf{H} \rangle$: $p_c = p_c(\langle \mathbf{H} \rangle)$. To the best of our knowledge, the dependence of the percolation threshold on the magnetic field was heuristically introduced in Ref. [38] for explaining the magnetic-field dependence of properties of MR fluids. In Ref. [10], it was proposed to fit the experimental data on the magnetorheological effect in MAEs by percolation-like dependences with the magnetic-field-dependent critical volume fraction $p_c$. In Ref. [39], we used the method of Padé approximants for estimating the concentration dependence of the effective elastic modulus of MAE for the two-dimensional case and obtained an analytical expression for $p_c(H)$. Recently, a strain-dependent percolation threshold in highly stretchable conductive composite materials has been reported [40].

Note that one can imagine situations when another mechanism of RS may be possible in composite materials, namely there is such a displacement of inclusions where local geometric anisotropy occurs. For example, when either inclusions themselves are elongated [41] or spherical particles are assembled into elongated agglomerates [42]. This mechanism deserves separate consideration. Calculation methods for such structures are reviewed in Ref. [5].

### III. Modification of the Bruggeman-Landauer approximation in the absence of RS

For calculating the effective coefficients, first we need to choose the approximation method. For example, when locally $\mathbf{D} = \varepsilon_0 \varepsilon(\mathbf{r}) \mathbf{E}$ is fulfilled. Then the effective coefficient (the effective dielectric constant) $\varepsilon^e$ is defined as in (1). The first phase has the relative permittivity $\varepsilon_1$, the second phase has the relative permittivity $\varepsilon_2$, with concentrations of phases being $p$ and $1-p$, respectively.

One of the most successful approximations for calculating the effective coefficients, working both in the case of weak and strong inhomogeneity, is the BL approximation [43,44], which is also known as the effective medium theory [45]. The expression for the effective dielectric constant is given by the equation

$$\frac{\varepsilon^e - \varepsilon_1}{\varepsilon_1 + 2\varepsilon^e} p + \frac{\varepsilon^e - \varepsilon_2}{\varepsilon_2 + 2\varepsilon^e}(1-p) = 0, \qquad (2)$$

or

$$\varepsilon^e = \frac{1}{4}\left\{(3p-1)\varepsilon_1 + (2-3p)\varepsilon_2 + \sqrt{\left[(3p-1)\varepsilon_1 + (2-3p)\varepsilon_2\right]^2 + 8\varepsilon_1\varepsilon_2}\right\}. \qquad (3)$$

In the case of a sufficiently large inhomogeneity (large ratio of $\varepsilon_1/\varepsilon_2$), with the growing concentration of inclusions $p$, a sharp increase in $\varepsilon^e$ occurs according to (3) at $p = p_c = 1/3$. We will call this concentration value the percolation threshold. In real composites, the value of $p_c$ depends on the manufacturing technology of the composite and can be significantly different from the value of 1/3 obtained in the conventional BL approximation. Therefore, the BL approximation (3) must be modified in such a way that the percolation threshold can take on any specified value. Such a modification was proposed in [46]. According to [46], equation (3) should be replaced by the following expression

$$\frac{\dfrac{\varepsilon^e - \varepsilon_1}{\varepsilon_1 + 2\varepsilon^e}}{1 + c(p,p_c)\dfrac{\varepsilon^e - \varepsilon_1}{\varepsilon_1 + 2\varepsilon^e}} p + \frac{\dfrac{\varepsilon^e - \varepsilon_2}{\varepsilon_2 + 2\varepsilon^e}}{1 + c(p,p_c)\dfrac{\varepsilon^e - \varepsilon_2}{\varepsilon_2 + 2\varepsilon^e}}(1-p) = 0, \qquad (4)$$

where the addends in the denominators proportional to $c(p,p_c)$ represent the generalization of the BL approximation by Sarychev and Vinogradov (SV). Specifically, the SV term is

$$c(p,p_c) = (1-3p_c)\left(\frac{p}{p_c}\right)^{p_c}\left(\frac{1-p}{1-p_c}\right)^{1-p_c}, \qquad (5)$$

where one can select $p_c$ in (5) according to the experimentally considered composite material. Explicitly, the expression for the effective dielectric constant, according to (4) and (5), can be written as

$$\varepsilon^e = \frac{1}{2(2+c)}\left\{(3p-1+c)\varepsilon_1 + (2-3p-c)\varepsilon_2 + \sqrt{\left[(3p-1+c)\varepsilon_1 + (2-3p-c)\varepsilon_2\right]^2 + 4\varepsilon_1\varepsilon_2(1-c)(2+c)}\right\}, \quad (6)$$

where, for the sake of simplicity, the arguments of $c(p, p_c)$ have been omitted.

Note that if $p_c = 1/3$ (percolation threshold in the BL approximation (2)-(3)), the SV term vanishes and the modified approximation (4) is transformed into (2). Appendix 1 explains the meaning of the term percolation threshold, which originates from the percolation theory, both in the context of the (original) BL approximation and in the SV modification of it.

How does the percolation threshold depend on the magnetic field? In [10], experimental data on the effective elastic moduli of MAE materials were analyzed for different concentration of iron particles and the following empirical relationship was proposed:

$$p_c(|\langle\mathbf{H}\rangle|) = p_c(0) e^{-\frac{(|\langle\mathbf{H}\rangle|)}{H_c}}, \quad (7)$$

where the value of the constant $H_c$ was determined to be equal to $H_c = 6.2 \cdot 10^5$ A/m. Accordingly to (7), the threshold $p_c$ decreases with increased $(|\langle\mathbf{H}(\mathbf{r})\rangle|)$, that, taken into account (5) and (6), means an increase in effective permittivity $\varepsilon^e$. Very recently, Puljiz *et al.* [47] experimentally demonstrated a virtual touching and detachment of rigid inclusions in a soft elastic matrix by application and withdrawing of a magnetic, which can be considered as an elementary step towards build-up (destruction) of an infinite cluster in magnetic field (or its removal).

## IV. Magnetodielectric effect in magnetoactive elastomers

Large theoretical effort has been devoted to the calculation of the effective dielectric constant of composites (see, for example, books [3,4,6]). When measurement is performed at a non-zero frequency and the dielectric has losses (associated with non-zero electrical conductivity), it is convenient to use complex variables to describe the properties of the phases. For a "bad" dielectric, the complex dielectric constant can be written as the sum [31] of two terms, the real part (lossless dielectric constant $\varepsilon'$) and the imaginary part $\varepsilon''$ associated with losses:

$$\hat{\varepsilon}(\omega) = \varepsilon' + i\varepsilon'' = \varepsilon' + i\frac{\sigma}{\omega\varepsilon_0}, \tag{8}$$

where $\sigma$ is the specific conductivity of a «non-ideal» dielectric, $\omega$ is the angular frequency, and $i$ is the imaginary unit. The apparent conductivity $\sigma$ can be recalculated into imaginary part of the relative permittivity: $\varepsilon'' = \sigma\omega^{-1}\varepsilon_0^{-1}$ [48]. In (8), we have used the physics convention where the harmonic time dependence has the form of $e^{-i\omega t}$.

Magnetodielectric effect is conventionally defined as the variation of the lossless permittivity $\varepsilon'$ due to externally applied magnetic fields [49]. To calculate the effective dielectric constant, we use the BL approximation (4) with SV terms, where now $\hat{\varepsilon}^e(\omega) = \varepsilon'^e + i\cdot(\sigma^e/\omega\varepsilon_0)$. Obviously, Eq. (4) should be now re-written for complex permittivities:

$$\frac{\dfrac{\hat{\varepsilon}^e - \hat{\varepsilon}_1}{\hat{\varepsilon}_1 + 2\hat{\varepsilon}^e}}{1 + c(p, p_c)\dfrac{\hat{\varepsilon}^e - \hat{\varepsilon}_1}{\hat{\varepsilon}_1 + 2\hat{\varepsilon}^e}} p + \frac{\dfrac{\hat{\varepsilon}^e - \hat{\varepsilon}_2}{\hat{\varepsilon}_2 + 2\hat{\varepsilon}^e}}{1 + c(p, p_c)\dfrac{\hat{\varepsilon}^e - \hat{\varepsilon}_2}{\hat{\varepsilon}_2 + 2\hat{\varepsilon}^e}} (1 - p) = 0. \tag{9}$$

Note that according to the RS model, the percolation threshold depends on the magnetic field (7), i.e. $c = c(p, p_c(|\langle \mathbf{H} \rangle|))$. This is the qualitative explanation of the magnetodielectric effect, because the dielectric

properties will depend on the external magnetic field as well. Figure 2 shows the dependence of the real part of the effective dielectric constant $\left(\operatorname{Re}\{\hat{\varepsilon}^e\} = \varepsilon'\right)$ as a function of the applied magnetic field.

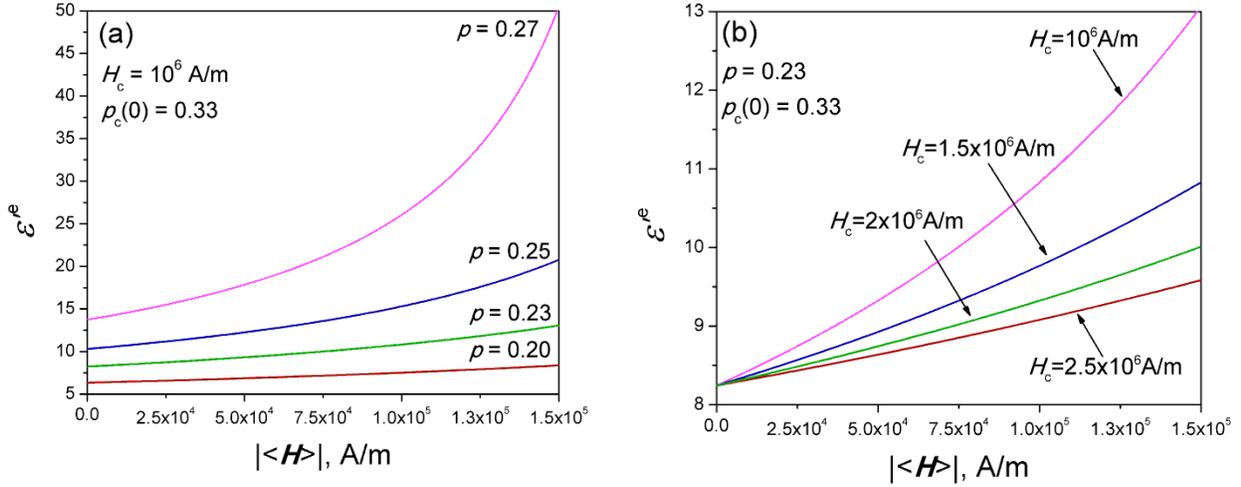

**Figure 2.** Field dependences of the effective lossless dielectric constant on the magnitude of the average magnetic field strength in the composite material. (a) Family of curves for a fixed value of the critical magnetic field $H_c$ and different volume particle concentrations. (b) Family of curves for a fixed particle concentration and different critical magnetic fields $H_c$. The following data for the constitutive materials at frequency $f$ of 1 kHz have been used: $\hat{\varepsilon}_1 = 1 + i1.8 \cdot 10^{14}$ (*i.e.*, good conductor, cf. [50]), $\hat{\varepsilon}_2 = 2.5 + i2.5 \cdot 10^{-3}$ [20,48].

Figure 2 (a) demonstrates the variation of the lossless dielectric constant increases with the increasing concentration of ferromagnetic particles. Figure 2 (b) shows that the variation of the lossless dielectric constant increases with decreasing critical field $H_c$. A selected value of $p_c(0)$ is close to 1/3 and agrees well with the value observed in similar materials [51]. It is rather obvious that the constant $H_c$ in (7) should depend on the stiffness of the matrix, i.e. the ability of inclusions to move under the influence of a magnetic field. The stiffer is the matrix, the larger constant $H_c$ must be expected. For the theoretically most rigid matrix, in which the particles practically do not move (the lower curve in Fig. 2 (b)), the effective dielectric constant is practically independent of the applied magnetic field. This is a direct indication of the relation of the magnetodielectric effect to the RS. The results of Figure 2 qualitatively agree well with experimental results obtained in Ref. [20] for different particle concentration and softness of the polymer matrix.

## V.  Effective magnetic permeability taking into account nonlinearity in the absence of RS

Let us consider the method of calculating the concentration dependence of the effective magnetic permeability in the absence of RS. This is the case in conventional polymer-based ferromagnetic composite materials [52-54] or in MAEs at low temperatures [11]. To be specific, consider a MAE specimen with rigidized PDMS polymer matrix at low temperatures (below 220 K) [11]. Such a sample is "frozen" and there is no displacement of inclusions when a magnetic field is applied to a specimen. However, it is impossible to use the BL approximation with the SV modification (4) directly. The local magnetic permeability of ferromagnetic inclusions, even in the hysteresis-free case, (nonlinearly) depends on the applied magnetic field. In turn, the magnetic field, in an inhomogeneous medium depends on coordinates in a complex way:

$$\mu_1 = \mu_1\left(\|\mathbf{H}(\mathbf{r})\|\right). \tag{10}$$

Thus, the calculation of the effective magnetic permeability, even in the absence of rearrangement of inclusions (i.e. RS), is nonlinear problem and the use of various approximate expressions for effective permeability, developed for linear problems, is not straightforward.

Various approximations have been proposed for calculating the effective coefficients in the case of nonlinearity of local kinetic coefficients. We shall follow the most suitable for our cases method proposed in [55-57], where a method of generalizing the approximations developed for linear problems to the nonlinear case was given. In the original two works [55-56], the power-type nonlinearities were considered, see also [58]. The effective properties of composites with fixed ferromagnetic inclusions were considered in [59-62]. The method proposed in [55,56] is briefly explained in Appendix 2. The effective magnetic permeability is found by solving a system of two nonlinear equations

$$\langle H^2 \rangle_1 = \frac{\langle \mathbf{H} \rangle^2}{p} \frac{\partial \mu^e(p, \tilde{\mu}_1, \mu_2)}{\partial \tilde{\mu}_1}, \qquad (11)$$

and

$$\mu^e(p, |\langle \mathbf{H} \rangle|) = \frac{1}{2(2+c(p,p_c))} \Big\{ (3p-1+c(p,p_c))\tilde{\mu}_1 + (2-3p-c(p,p_c))\mu_2 + \\ + \sqrt{[(3p-1+c(p,p_c))\tilde{\mu}_1 + (2-3p-c(p,p_c))\mu_2]^2 + 4\tilde{\mu}_1\mu_2(1-c(p,p_c))(2+c(p,p_c))} \Big\}, \qquad (12)$$

where $\tilde{\mu}_1 = \mu_1\left(\sqrt{\langle H^2 \rangle_1}\right)$ and the subscript in $\langle ... \rangle_1$ means averaging over the first phase (magnetic inclusions).

To compare the theoretical model obtained by solving the system of equations (11),(12) with the experimental data, it is necessary to choose the explicit dependence of the magnetic permeability of inclusions on the magnetic field. As a rule, for micrometer-sized carbonyl iron particles, the empirical Fröhlich-Kennelly formula [63] is employed:

$$\mu_1(H) = \frac{\mu_{ini} + (\mu_{ini} - 1)\frac{H}{M_s}}{1 + (\mu_{ini} - 1)\frac{H}{M_s}}. \qquad (13)$$

Here we use the dependence of the magnetic permeability of inclusions in the form (12). The value of the two parameters that determine the numerical values are taken from the paper [64]: $\mu_{ini} = 132$, and $M_s = 1.99 \cdot 10^6$ A/m. Figure 3 shows the field and concentration dependences of the magnetic permeability of MAE obtained from the solution of the system of equations (11-12) and a comparison with the experiment of [11]. It is seen that there is an excellent agreement between theory and experiment for a conventional set of parameters.

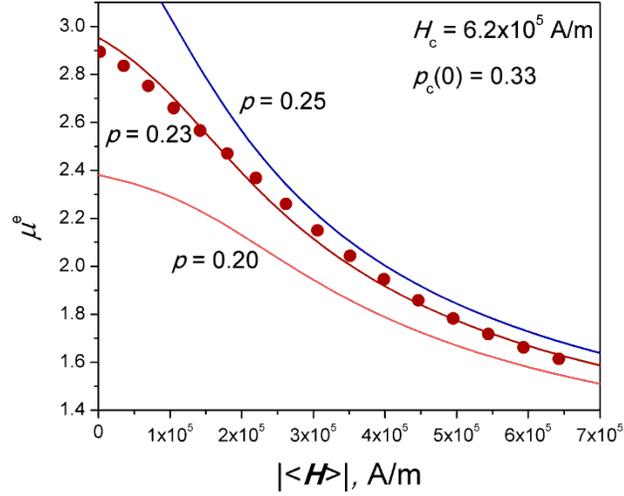

**Figure 3.** Field dependences of the effective magnetic permeability of a MAE sample at $T = 150$ K for different particle concentrations. The filled circles denote experimental values, derived from magnetization curves of Ref. [11].

As expected, the larger the concentration of inclusions, the larger the effective magnetic permeability (for the same values of the applied magnetic field). The magnetic permeability of the inclusions $\mu_1(|\langle \mathbf{H} \rangle|)$ decreases with increasing magnetic field strength $|\langle \mathbf{H} \rangle|$ according to (12), and so does $\mu^e(|\langle \mathbf{H} \rangle|)$. Note that, as follows from general considerations, in the absence of RS, the dependences of the effective magnetic permeability are monotonic (declining) functions of $|\langle \mathbf{H} \rangle|$.

## VI. Effective properties taking into account both nonlinearity and RS: effective magnetic permeability of MAE at room temperature

As above, we describe RS as a change in the percolation threshold in a magnetic field (7). To calculate the effective properties, the nonlinear system of equations (11)-(12) must be modified again. Now, the percolation threshold included in the SV term (5) depends on magnetic field: $p_c = p_c(|\langle \mathbf{H} \rangle|)$. Thus, equation (12) takes the following form

$$\mu^e(p,|\langle \mathbf{H}\rangle|) = \frac{1}{2(2+c(p,p_c(|\langle \mathbf{H}\rangle|)))} \{[3p-1+c(p,p_c(|\langle \mathbf{H}\rangle|))]\tilde{\mu}_1 + [2-3p-c(p,p_c(|\langle \mathbf{H}\rangle|))]\mu_2 +$$

$$+([[3p-1+c(p,p_c(|\langle \mathbf{H}\rangle|))]\tilde{\mu}_1 + [2-3p-c(p,p_c(|\langle \mathbf{H}\rangle|))]\mu_2]^2 +$$

$$+4\tilde{\mu}_1\mu_2[1-c(p,p_c(|\langle \mathbf{H}\rangle|))][2+c(p,p_c(|\langle \mathbf{H}\rangle|))])^{1/2}\}. \qquad (14)$$

Figure 4 shows the results of the calculation of the effective magnetic permeability according to the modified BL equation taking into account the nonlinearity and dependence of the percolation threshold on the magnetic field and the experimental dependences obtained in [11,18].

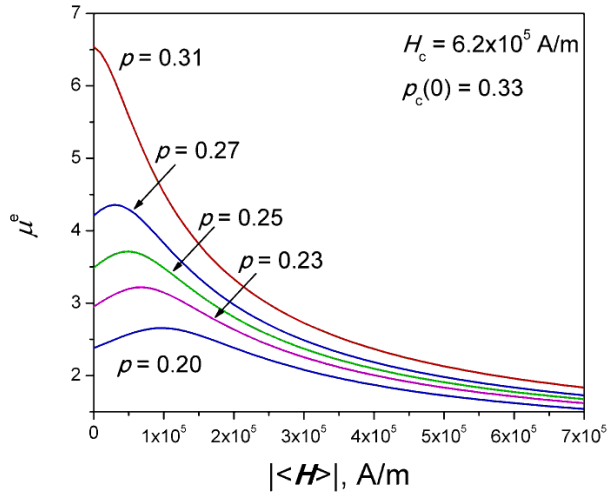

**Figure 4.** Field dependence of the effective magnetic permeability for different particle concentrations $p$.

In Fig. 4, the field dependences of the effective magnetic permeability are fundamentally different from those in Fig. 3, which are obtained in the absence of RS. In the case of RS, the dependence of magnetic permeability on the applied magnetic field ceases to be monotonous. It is interesting to note that the higher the concentration of inclusions, the weaker the field, where the maximum of permeability occurs. The manifestation of RS on the field dependence of magnetic permeability is most clearly seen in Fig. 5, which compares the experimental curve at room temperature (when RS occurs) [11] and two theoretical dependences, first is with RS, the other is without it. All dependencies are obtained for the same concentration of inclusions $p = 0.23$.

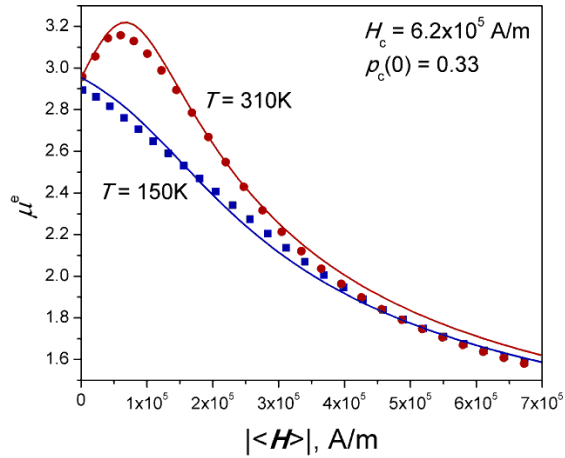

**Figure 5.** Field dependence of the effective magnetic permeability for two fundamentally different cases. The red curve shows the calculated field dependence of the effective magnetic permeability, obtained from the system of nonlinear equations (8, 9) in the presence of RS (room temperature), The green curve presents the calculated field dependence of the effective magnetic permeability in the absence of RS (low temperature). The filled circles denote experimental values, derived from magnetization curves of Ref. [11].

### VII. Discussion

Despite the fact that a composite is far from the percolation threshold and it is not described by percolation theory, it is important that effective coefficients do not depend on the actual concentration of inclusions, but on the proximity of the concentration to the percolation threshold. Percolation theory implies that effective coefficients are characterized by critical indexes, the concentration of inclusions is close to the percolation threshold and the ratio of local material property of phases, for example, the electrical conductivity is very large. Real composites possessing different microstructures (which results in a specific percolation threshold) with the same concentration of inclusions will lead to different values of the effective coefficients. This can be related to the fact that the concentration of inclusions in some of the composites is nearer to the percolation threshold.

The RS concept allows one to explain experimental dependences of different physical quantities on an applied magnetic field in MAEs. First is the magnetodielectric effect, which is the significant increase of the effective dielectric constant in applied magnetic fields. The second effect is a qualitatively different behavior of the dependence of effective magnetic permeability on the magnitude of the magnetic field in MAE at room and nitrogen temperatures. At low temperatures, according to experimental data [11], the effective magnetic permeability decreases with increasing field. This is due to a decrease in the magnetic permeability of inclusions with increasing field (13). At room temperatures, local maxima on a generally descending dependence of $\mu^e$ on the increasing magnetic field strength $H$ (see Fig. 4) have been experimentally observed in Ref. [11]. Qualitatively similar result has been reported before in Ref. [17]. Of course, this maximum is related to RS. As the field increases, the PC grows, which is interpreted as a decrease in the percolation threshold and, accordingly, the increase in $\mu^e$. On the other hand, since the local magnetic permeability of inclusions decreases with an increase in the magnetic field (13), growth of the magnetic field contributes to a decrease in the effective magnetic permeability. The presence of two opposite trends can lead to the appearance of a maximum in the dependence of the effective magnetic permeability on the external field. To summarize, in this paper, a generalized theoretical approach for approximate calculation of effective properties of particulate composites with RS of the filling particles is formulated. It is employed for alternative explanation of two different physical effects in MAEs. As far as the magnetorheological (or field-stiffening) effect in MAEs is concerned, in the framework of the suggested approach, one has to incorporate the movable percolation threshold into the BL system of equations of elasticity. To the best of our knowledge, such a modification is not available at the present stage of research.

## VIII. Conclusion

The main principal points of our model are:

1. The properties of MAE-type composite materials are due to particle displacements (rearrangement of inclusions) in a magnetic field, i.e. changes in the microstructure of the filling. We referred to this physical phenomenon as restructuring. We emphasize that the concentration of inclusions in the sample does not change with RS, only their concentration in the vicinity of the pre-cluster changes.

2. In the theoretical description of composites with different values of percolation threshold and RS, it is reasonable to speak of a change in the "distance" to the percolation threshold. So far, the only logical way to describe the properties of composites with RS is to introduce a variable percolation threshold.

3. Standard equations that allow one to calculate effective coefficients, for example, the Maxwell, Maxwell-Garnett, Clausius-Mossotti, Bruggeman-Landauer approximations, etc., are not capable of describing a composite material with RS, because they describe a composite with a constant, given microstructure (distribution of inclusions). For example, the conventional Bruggeman-Landauer approximation in the three-dimensional case always(!) gives a percolation threshold approximately equal to 1/3 (the concentration value at which the maximum change in the concentration dependence of the effective coefficient occurs).

4. It is necessary to adapt the calculation methods for the effective coefficients in such a way that they describe composites with a variable percolation threshold. For this purpose, we employed the BL approximation, in which we introduced the SV modification. However, unlike the SV modification of the BL approximation, we consider the modified percolation threshold to be dependent on external conditions, in our case, on the applied magnetic field.

**Appendix 1.**

**Percolation threshold in the framework of the Bruggeman-Landauer approximation**

Consider a two-phase composite with conductivities of phases $\sigma_1 > \sigma_2$. According to the mean-field approximation (BL approximation), the effective conductivity of the medium with concentration $p$ of the first phase is determined by solving the system of equations (2), (3) (where the dielectric constants must be simply replaced by conductivities). For infinitely large inhomogeneity, when the ratio of the specific conductivities of the phases tends to infinity: $((\sigma_1/\sigma_2) \to \infty \Leftrightarrow (\sigma_2/\sigma_1) \to 0)$, a sharp change in the concentration behavior of the effective conductivity $\sigma^e$ occurs at a concentration $p$ of 1/3. Note that the realization of $\sigma_1/\sigma_2 \to \infty$ is possible in two limits. In the first limit, $\sigma_1$ has a finite value ($\sigma_1 \neq \infty$), $\sigma_2$ goes to zero ($\sigma_2 \to 0$). In this case, $\sigma^e$ has a non-zero value only for concentrations $p > 1/3$. In the second limit, $\sigma_2$ has a finite value ($\sigma_2 \neq 0$), $\sigma_1$ goes to infinity ($\sigma_2 \to \infty$) and the finite value $\sigma^e \neq \infty$ occurs for concentrations $p < 1/3$, with the effective conductivity $\sigma^e$ being equal to infinity at $p > 1/3$. The value of 1/3 obtained in the BL approximation is called the percolation threshold.

According to percolation theory [3], percolation threshold is the value of concentration when a so-called infinite cluster (connected path in the first phase through the entire system) occurs. When studying physical processes, for example, conductivity, the formation of a connected path leads to a sharp change of the effective coefficients. The numerical value of the percolation threshold is not universal and is different for different types of lattices and structures (for example, permeable and impermeable inclusions). Experimentally, for different composites, different values of percolation thresholds are observed.

Let us now consider the case of a finite heterogeneity, for example, a finite ratio of conductivities: $(\sigma_1/\sigma_2) \neq \infty \Leftrightarrow (\sigma_2/\sigma_1) \neq 0$. As far as the calculation of effective conductivity is concerned, the abrupt transition (kink) of the behavior of effective conductivity becomes smooth, cf. Figure 6. It is reasonable to call the concentration value $p_c^\sigma$ at which the maximum variation in the behavior of $\sigma^e$ occurs, the percolation threshold. The upper index in $p_c^\sigma$ indicates that this parameter does not "originate" from the infinite-cluster problem, but from the BL approximation. When $\sigma_1/\sigma_2 \to \infty$, the percolation threshold $p_c^\sigma$ becomes equal to 1/3. With a finite ratio $\sigma_1/\sigma_2$, the maximum change in concentration behavior of $\sigma^e$ corresponds to the root of the equation $d^3\sigma^e/dp^3 = 0$, see Fig. 6. According to (3)

$$\frac{d^3\sigma^e}{dp^3} = 162 \frac{\sigma_1\sigma_2(\sigma_1-\sigma_2)^3(\sigma_1-2\sigma_2-3p\sigma_1+3p\sigma_2)}{\left[(\sigma_1-2\sigma_2-3p\sigma_1+3p\sigma_2)^2+8\sigma_1\sigma_2\right]^{5/2}}. \tag{15}$$

By equating the numerator of the above formula to zero, we find the dependence of $p_c^\sigma$ on the ratio $\sigma_2/\sigma_1$, see Fig. 7 (a):

$$p_c^\sigma = \frac{1}{3} \cdot \frac{1-2\sigma_2/\sigma_1}{1-\sigma_2/\sigma_1}, \quad \sigma_1 > \sigma_2, \quad p_c^\sigma\left(\frac{\sigma_2}{\sigma_1} \to 0\right) \to \frac{1}{3}. \tag{16}$$

The condition $\sigma_1 > \sigma_2$ in (2) means that we are looking for the percolation threshold for conductivity that corresponds to the appearance of an infinite cluster in the first phase. As can be seen from Figure 7 (a), with nonzero but not very small ratio $\sigma_2/\sigma_1 < 0.1$, the difference of $p_c^\sigma$ from $p_c$ = 1/3 is not significant. There is also a percolation threshold $p_c^\sigma$ for the second phase, this means the concentration of the first phase at which an infinite cluster in the second phase occurs, see Fig. 7 (a):

$$p_c^\sigma = \frac{1}{3} \cdot \frac{\sigma_1/\sigma_2 - 2}{\sigma_1/\sigma_2 - 1}, \quad \sigma_1 < \sigma_2, \quad p_c^\sigma\left(\frac{\sigma_1}{\sigma_2} \to 0\right) \to \frac{2}{3}. \tag{17}$$

As can be seen from Figure 7 (a), with nonzero but not very small ratio $\sigma_1/\sigma_2 < 0.1$, the difference of $p_c^\sigma$ from 2/3 is not significant.

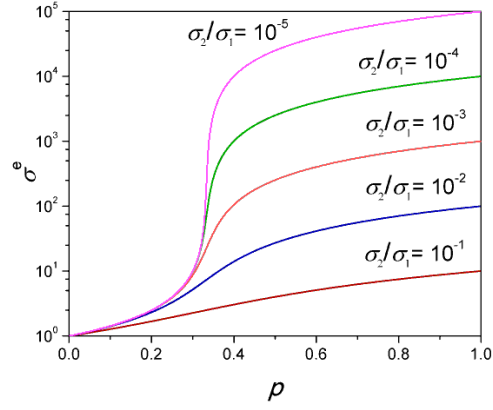

**Figure 6.** Dependence of the effective conductivity $\sigma^e$ on the concentration $p$ of the first phase for different ratios $\sigma_2/\sigma_1$.

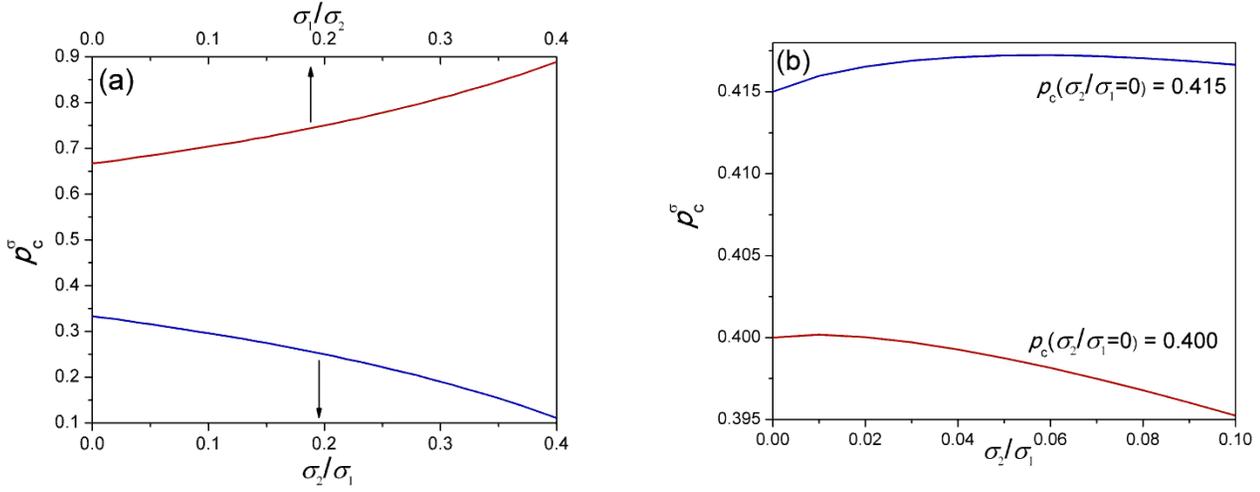

**Figure 7.** (a) Dependence of the percolation threshold $p_c^\sigma$ on the ratio $\sigma_2/\sigma_1$ (lower horizontal axis) or $\sigma_1/\sigma_2$ (upper horizontal axis) in the conventional BL approximation. (b) Dependence of the percolation threshold $p_c^\sigma$ on the ratio $\sigma_2/\sigma_1$ in the modified BL approximation for two different values of $p_c^\sigma((\sigma_2/\sigma_1)=0) \neq 1/3$.

Let us consider how $p_c^\sigma$ is affected when the SV modification is introduced into the BL approximation. The BL-equation is now replaced by (4). For $\sigma_2/\sigma_1 \to 0$, the percolation threshold $p_c^\sigma$ goes to the limiting value specified in the SV modification (4), where it is designated as $p_c$. In the case of a finite ratio $\sigma_2/\sigma_1 \neq 0$, the value of $p_c^\sigma$ differs from the given value of $p_c$ in (4). As can be seen from Figure 7 (b), for a small value of the ratio $\sigma_2/\sigma_1 < 0.1$, such a deviation is insignificant.

**Appendix 2.**

**Nonlinear equations for the effective magnetic permeability**

For any point belonging to the first phase, a local relation $\mathbf{B} = \mu_0 \mu_1(|\mathbf{H}|)\mathbf{H}$ holds. According to the approximation of the mean-field theory, in [55,56] this relation is replaced by $\mathbf{B} = \langle \mu_1(|\mathbf{H}|) \rangle_1 \mu_0 \mathbf{H}$, where the subscript 1 means averaging over the volume of the first phase. Further, a replacement $\langle \mu_1(|\mathbf{H}|) \rangle_1 \to \mu_1\left(\sqrt{\langle H^2 \rangle_1}\right)$ is made to enable the use of a self-consistent approximation. Thus, the composite is now a two-phase medium, each of which is characterized by constants $\tilde{\mu}_1 = \mu_1\left(\sqrt{\langle H^2 \rangle_1}\right)$ and $\mu_2$. We emphasize that now, in contrast to $\mu_1(|\mathbf{H}|)$, $\tilde{\mu}_1$ does not depend on the coordinates. To determine the average square of the field in the first phase, a decoupling scheme is employed (cf. [65], for details see [5,66]). For a sample with homogeneous boundary conditions at infinity ($\mathbf{B}(\mathbf{r} \to \infty) = const$, $\mathbf{H}(\mathbf{r} \to \infty) = const$), the volume-averaged product of the fields is decoupled: $\langle \mathbf{BH} \rangle = \langle \mathbf{B} \rangle \langle \mathbf{H} \rangle$. Then, on the one hand $\langle \mathbf{BH} \rangle / \mu_0 = \langle \mu H^2 \rangle = p \tilde{\mu}_1 \langle H^2 \rangle_1 + \mu_2 (1-p) \langle H^2 \rangle_2$, and on the other hand $\langle \mathbf{B} \rangle \langle \mathbf{H} \rangle / \mu_0 = \mu^e \langle \mathbf{H} \rangle^2$ are fulfilled. Thus, $\mu^e \langle \mathbf{H} \rangle^2 = p \tilde{\mu}_1 \langle H^2 \rangle_1 + \mu_2 (1-p) \langle H^2 \rangle_2$, from which it the immediately follows that

$$\langle H^2 \rangle_1 = \frac{\langle \mathbf{H} \rangle^2}{p} \frac{\partial \mu^e(p, \tilde{\mu}_1, \mu_2)}{\partial \tilde{\mu}_1}, \qquad (18)$$

For $\mu^e(p, |\langle \mathbf{H} \rangle|)$, the BL formula now has the following form:

$$\mu^e(p, |\langle \mathbf{H} \rangle|) = \frac{1}{2(2+c(p,p_c))} \{(3p-1+c(p,p_c))\tilde{\mu}_1 + (2-3p-c(p,p_c))\mu_2 +$$
$$+ \sqrt{[(3p-1+c(p,p_c))\tilde{\mu}_1 + (2-3p-c(p,p_c))\mu_2]^2 + 4\tilde{\mu}_1 \mu_2 (1-c(p,p_c))(2+c(p,p_c))} \}. \qquad (19)$$

Thus, after choosing a particular expression for the effective coefficient of a two-phase medium with constant coefficients ($\mu^e = \mu^e(p, \mu_1, \mu_2)$), we obtain from (18) and (19) a system of two nonlinear equations for determining $\langle H^2 \rangle_1$ and $\mu^e$.

**Acknowledgement**

We thank Professor Tetsu Mitsumata for experimental details of Ref. 10. We are grateful to Dr. Thomas Gundermann for friendly communications with respect to Figure 5 (b) of Ref. 37. The authors thank the Bavarian Academic Center for Central, Eastern and Southeastern Europe (BAYHOST) for financial support of the reciprocal visits (grant No. MB-2018-2/5). The work of M.S. was funded by the Deutsche Forschungsgemeinschaft (DFG, German Research Foundation) – project number 389008375.